\RequirePackage{fixltx2e}
\documentclass[a4paper]{isp}
\usepackage[american]{babel}
\usepackage{graphicx}
\usepackage{amsmath,amsthm,mathtools}
\usepackage{url}
\usepackage[capitalise,nameinlink]{cleveref}
\usepackage{xcolor}
\def\aap{A\&A\,  }

\def\aj{AJ  }
\def\apj{ApJ\,  }
\def\apjl{ApJ Letters,  }
\def\apss{Astrophysics and Space Science  }

\def\mnras{MNRAS\,  }

\def\snr1993j{SN\,1993J~}
\def\grb{GRB 050814~}
\def\rgb2{GRB 060729~}
\begin{document}
\pdfgentounicode=1
\title
{
Energy Conservation in the thin  layer approximation:
IV. The light curve for supernovae 
}
\author{Lorenzo Zaninetti}
\institute{
Physics Department,
 via P. Giuria 1, I-10125 Turin, Italy \\
 \email{zaninetti@ph.unito.it}
}

\maketitle

\begin {abstract}
The  light curves (LC) for
Supernova  (SN) 
can be modeled adopting  
the conversion of the flux of kinetic energy into radiation.
This conversion requires  an analytical or a numerical
law of motion for the expanding radius of the SN.
In the framework of conservation of energy 
for the  thin layer approximation 
we  present 
a classical trajectory based on 
a power law profile for the density,
a relativistic trajectory based 
on the Navarro--Frenk--White profile for the density,
and 
a relativistic trajectory based on
a power law behaviour for the swept mass.
A detailed simulation of the LC requires 
the evaluation of the optical depth as a function of time.
We modeled the LC of SN~1993J in different
astronomical bands,
the LC  of GRB 050814 
and
the LC GRB 060729 in the keV region.
The time  dependence of the magnetic field of equipartition
is derived from the theoretical formula for the luminosity.
\end{abstract}

\section{Introduction}

The number of observational and theoretical analyses  of the light curves
(LCs)
 for supernovae 
(SN) has increased in recent years.
We list    some of the recent treatments.
The LC of the
type Ia supernova  2018oh 
has an  unusual two-component shape \cite{Dimitriadis2019},
the radio LC of  SN 1998bw   shows a
        double-peak profile, possibly associated with density variations
        in the circumstellar medium  \cite{Palliyaguru2019},
the R-band LCs of 265 SNs
from the Palomar Transient  Factory
were followed  and 
a model-independent LC template was  built from  this 
data-set  \cite{Papadogiannakis2019},
SN 2007D (which is a  luminous type Ic supernova)
has  a narrow  LC and high peak luminosity
that were explored with a multi-band model \cite{Wang2019},
evolutionary models for the LC were introduced
using the  
STELLA software application \cite{Ricks2019},
the conversion  of the
kinetic energy of ejecta to radiation at the reverse
and forward shocks was introduced in \cite{Tsuna2019},
the LC was modeled  
in the framework of the radioactive decay 
of  $^{56}$Co, $^{57}$Co and $^{55}$Fe \cite{Kushnir2020},
the cosmological importance of the LC was analysed 
by  \cite{Koo2020},
and PS15dpn is a luminous rapidly rising Type Ibn SN 
which was modeled in the framework of 
the circumstellar interaction (CSI) model plus        
$^{56}$Ni decay \cite{Wang2020}. 
The previous papers  leave  a series of
questions unanswered.
\begin{itemize}
\item
Given the observational fact that the
radius--time
relation  in young SNRs follows a power law,
is it possible to
find a theoretical law of motion 
in the framework of the classical energy conservation?
\item
Can we express the flux of kinetic energy 
in  an analytical  way
in  a medium which is 
characterized by a decreasing density?
\item 
Can we parametrize
the conversion of the analytical or numerical
flux of kinetic energy into  the  observed
luminosity?
\item
Can we model the double-peak profile for the LC
in the framework of the temporal variations
of the optical thickness?
\item
Can we apply the classical and relativistic approaches to the
LC of SNs and  Gamma Ray Bursts (GRBs)?
\item 
Can we model   the  evolution of the
magnetic field? 
\end{itemize}
This  paper is structured as follows.
In Section \ref{section_preliminaries}
we explore the power law  fit model.
Section \ref{section_luminosity} 
reviews the classical and relativistic 
conversion of the flux of kinetic energy into 
luminosity.
Section \ref{section_classical} presents
some  analytical results  for 
a classical law of motion,
Section \ref{section_relativistic}
introduces  two  new relativistic equations of motion,
Section \ref{section_astrophysical} presents
the simulation of the LC for one SN and two GRBs
and
Section \ref{section_magnetic} presents 
the temporal evolution of the magnetic 
field as well some evaluations 
for 
 the accelerating clouds 
due to the Fermi II acceleration mechanism.

\section{Preliminaries}

\label{section_preliminaries}

This section  presents the analysed SN 
and GRB,
introduces the adopted statistics, 
and 
reviews the power law model as a useful fit
for the radius--time relation in SNs.

\subsection{The analysed SN and GRB}

The first  SN to be analysed is  
\snr1993j,  
for  which the  temporary  radius of expansion
has been measured for  $\approx$ 10 yr in
the radio band \cite{Marcaide2009,Zaninetti2014a}.
Here we processed for  the case of
\snr1993j the LC   for  the $R$ band 
as reported in Figure 5 in  \cite{Zhang2004}, 
the $V$ band for a short number of 
days, $\approx$ 63 days, which shows an oscillating behaviour, 
see Figure 4 
in \cite{Benson1994},
the luminosity of the $H-\alpha$ plotted
with the 2.0--8.0 keV LC as reported in Figure 5 in  \cite{Chandra2009}
and
the radio flux  density  at 15.2 GHz as 
observed by the Ryle Telescope \cite{Pooley1993}
with data available at \url{http://www.mrao.cam.ac.uk/~dag/sn1993j.html}.

The second  object to be    analysed  is  \grb 
at 0.3--10 keV, which covers the time interval 
[$10^{-5}-3$] days,
see \cite{Jakobsson2006}
with data available at  \url{https://www.swift.ac.uk/xrt_live_cat/150314}.

The third  object to be    analysed  is  \rgb2
observed 
by the Ultraviolet and Optical Telescope (UVOT) 
in the time interval 
[$10^{-2}-26$] days,
see Figure 1 in \cite{Cano2011}.

\subsection{The statistics}

The adopted  statistical parameters  are 
the percent error, $\delta$, between
the theoretical value and approximate value,
and  
the merit function $\chi^2$ evaluated
as
\begin{equation} 
\chi^2 =\sum_{i=1}^N \Big 
[ \frac{ y_{i,theo}-y_{i,obs} }{ \sigma_i } \Big ]^2
\label{chiquaredef}
\end{equation}
where $y_{i,obs}$ and $\sigma_i$   represent
the observed value and its error at position $i$,
$y_{i,theo}$ is  the theoretical value at position $i$
and  $N$ is the number of elements of the sample.

\subsection{The power law model}

The equation for   the expansion  of a SN
may   be modeled by a power law  
\begin{equation}
r(t) = C{t}^{\alpha_{{{\it fit}}}}
\label{rpower}
\quad ,
\end{equation}
where
$r$ is the radius  of the expansion,
$t$ is the time,
and  $\alpha_{\it fit}$ is an exponent which  can be found 
numerically.
The velocity is
\begin{equation}
v(t) = 
C{t}^{\alpha_{{{\it fit}}}-1}\alpha_{{{\it fit}}}
\quad .
\label {vpower}
\end{equation}
As a practical   example,  the   radius (pc)  time (yr)  relation 
in \snr1993j is 
\begin{equation}
r(t) = 0.0155 \times t^{0.828} \,pc 
\quad  ,
\end{equation}
when  $0.49\,yr <t < 10.58 \, yr $, 
see also Table \ref{datafitsnr1993j}.

\section{Luminosity}

\label{section_luminosity}

In these subsections we analyse the 
classical and relativistic conversion
of the flux of kinetic energy into luminosity.
The absorption of the produced radiation
is parametrized by the optical thickness.

\subsection{Conversion of energy}

In the {\it classical} case, the   
rate of transfer of
mechanical energy, $L_{m}$, is
\begin{equation}
L_{m}(t) = \frac{1}{2}\rho (t)4 \pi r(t)^2 v(t)^3
\quad ,
\end{equation}
where $\rho(t)$, $r(t)$ and  $v(t)$ are the temporary  
density, radius and velocity of the SN.
We assume    that the density in front of the
advancing  expansion scales as
\begin{equation}
\rho(t) = \rho_0 ( \frac{r_0}{r(t)} ) ^{d}
\quad ,
\end{equation}
where $r_0$ is the radius at $t_0$ 
and  $d$ is a parameter  which allows
matching the observations; as an example,  
a value of $d=3$ is reported in
\cite{Nagy2014}.
With the above assumption,
the mechanical luminosity is  
\begin{equation}
L_{m}(t) = \frac{1}{2}   
\rho_0 ( \frac{r_0}{r(t)} ) ^{d}
 4 \pi r(t)^2 v(t)^3
\quad .
\label{eqnlmtrho}
\end{equation}

The    mechanical luminosity  in the case of  a power law 
dependence for  the
radius is 
\begin{equation}
L_{m}(t) = 2\,\rho_{{0}}{r_{{0}}}^{d}{C_{{{\it fit}}}}^{-d+5}{t}^{-3+ \left(
-d+5
 \right) \alpha_{{{\it fit}}}}\pi\,{\alpha_{{{\it fit}}}}^{3}
\quad .
\label{lmalpha}
\end{equation}

The energy  fraction  of
the mechanical luminosity
deposited
in the frequency $\nu$, $L_{\nu}$, 
is assumed to be proportional
to the mechanical luminosity through a constant
$\epsilon_{\nu}$
\begin{equation}
L_{\nu} = \epsilon_{\nu}  L_m
\quad .
\end{equation}
The flux  at  frequency  $\nu$ and  distance  $D$
is
\begin{equation}
F_{\nu} = \frac{\epsilon_{\nu}  L_m}{4 \pi D^2}
\quad .
\label{flux_classical}
\end{equation}
For  practical   purposes,  we impose   a match between 
the observed    luminosity,  $L_{obs}$, and
the theoretical luminosity,  $L_m$, 
\begin{equation}
L_{obs} = C_{obs}  L_m
\label{lcobs}
\quad ,
\end{equation}
where  $C_{obs}$  is a constant which equalizes the observed and 
the theoretical luminosity and varies on the base 
of the selected astronomical band.
In a analogous way,  the observed absolute magnitude is
\begin{equation}
M_{obs} = -\log_{10} (L_m) + k_{obs} 
\label{mobs}
\quad ,
\end{equation}
where $k_{obs}$ is a constant.
In the {\it relativistic}case 
the rate of transfer  
of
mechanical energy, $L_{m,r}$, 
assuming the same scaling for the density in the advancing layer, 
is
\begin{equation}
L_{m,r}(t) =
4\,{\frac {\pi\,{r(t)}^{2}\rho_{{0}}{c}^{3}\beta(t)}{1-{\beta(t)}^{2}} \left( 
{\frac {r_{{0}}}{r}} \right) ^{d}}
\quad ,
\label{lumrel}
\end{equation}
where $\beta(t)=\frac{v(t)}{c}$,
for more details, see \cite{Zaninetti2015c}.

A useful formula is that for the minimum magnetic field
density, $B_{min}$,
\begin{equation}
B_{min} = 1.8 (\eta \frac{L_{\nu}}{V})^{2/7} \nu^{1/7}\quad T
\quad ,
\end{equation}
where $\nu$  is  the considered frequency of synchrotron emission,
$L_{\nu}$ is the luminosity of the radio source at   $\nu$,
$V$ is  the volume
involved, and  
$\eta= \frac{\epsilon_{total}}{\epsilon_e}$ is a constant which connects
the relativistic  energy of the electrons, $\epsilon_e$,
with the total energy in non-thermal phenomena, 
$\epsilon_{total}$, 
see formula (16.50) in \cite{Longair2011}
or    
formula (7.14) in \cite{Pacholczyk1970}.

\subsection{Absorption}

The presence  of  the absorption
can be parametrized
introducing a
slab of optical thickness $\tau_{\nu}$.
The emergent  intensity $I_{\nu}$ after
the entire slab is
\begin{equation}
I_{\nu} = \int_0^{\tau_{\nu}} S_{\nu} e^{-t} dt
\quad ,
\end{equation}
where  $S_{\nu}$ is a uniform source function.
Integration gives
\begin{equation}
I_{\nu} = S_{\nu} (1-e^{-\tau_{\nu}})
\quad ,
\end{equation}
see formula 1.30 in \cite{rybicki}.
In the case  of an optically  thin medium,
$\tau_{\nu}=\infty$, 
the observed   luminosity can be derived
with Equation (\ref{lcobs}), but
otherwise, 
the following equation should be used:
\begin{equation}
L_{obs} = C_{obs}\,  L_m \, (1-e^{-\tau_{\nu}})
\label{lcobstau}
\quad ,
\end{equation}
where  $\tau_{\nu}$ is a function of time.
For the case of the apparent  magnitude,
we have
\begin{equation}
m_{obs} = -\log_{10} (L_m)-\log_{10}(1-e^{-\tau_{\nu}})  + k_{obs} 
\label{mobstau}
\quad .
\end{equation}
The value of $\tau_{\nu}$ can be derived with the following equation:
\begin{equation}
\tau_{\nu} =
-\ln  \left(1 -{{\rm e}^{- \left( m_{{{\it obs}}}-m_{{{\it theo}}}
 \right) \ln  \left( 10 \right) }} \right) 
\end{equation}
where $m_{theo}$ and $m_{obs}$  represent the theoretical 
and the observed apparent magnitude.
Due to the complexity of the time dependence 
of $\tau_{\nu}$, a polynomial approximation
of degree $M$
is used:
\begin{equation}
\tau_{\nu}(t)  =
  a_0
+ a_1 \,t
+ a_2 \,t^2
+ \cdots
+ a_M\, t^M  
\quad  ,
\label{polynomial}
\end{equation}
with more details in \cite{press}.
In some cases  we  apply the logarithms
to the  pair of data, i.e. $\log_{10}({x_i})$ and 
$\log_10({y_i})$; we call this the {\it logarithmic 
polynomial approximation}.

The absorption  in the relativistic  case  is assumed 
to be the same  once the classical luminosity, $L_m$,
is replaced by the  relativistic  luminosity 
$L_{m,r}$  
\begin{equation}
L_{obs} = C_{obs}\,  L_{m,r} \, (1-e^{-\tau_{\nu}})
\label{lcobstaurel}
\quad ,
\end{equation}
and 
\begin{equation}
m_{obs} = -\log_{10} (L_{m,r})-\log_{10}(1-e^{-\tau_{\nu}})  + k_{obs} 
\label{mobstaurel}
\quad .
\end{equation}

\section{A classical equation of motion}

\label{section_classical}
\label{motionpowerlaw}
\begin{figure*}
\begin{center}
\includegraphics[width=7cm,angle=-90]{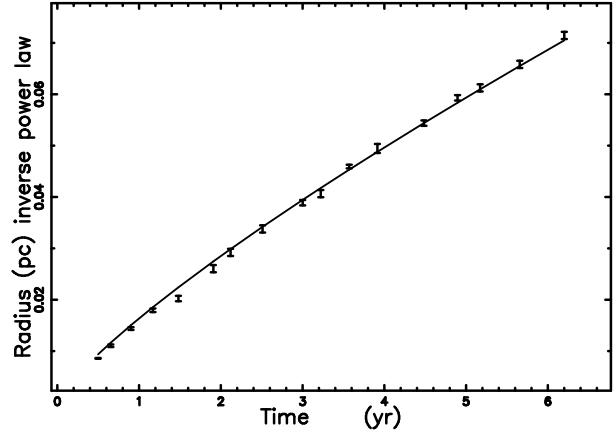}
\end {center}
\caption
{
Theoretical radius as given
by Eq.~(\ref{traiettoriapotenza}),
$v_0=4000 \,\frac{km}{s}$,
$t_0=10 \, yr$ 
and  $t=5\,10^4 yr$.
The model is the conservation of the classical energy
in the presence of an inverse power law  profile for the density.
}
\label{trajpowerlaw}
    \end{figure*}

Let us  analyse the  case  of conservation of energy 
in the thin layer approximation 
in the presence of a power law profile of density 
of the type
\begin{equation}
 \rho (r;r_0)  = \{ \begin{array}{ll}
            \rho_c                      & \mbox {if $r \leq r_0 $ } \\
            \rho_c (\frac{r_0}{r})^{\alpha}    & \mbox {if $r >     r_0 $.}
            \end{array}
\label{piecewisealpha},
\end{equation}
where 
$\rho_c$ is the density at $r=0$,
$r_0$ is the radius after which the density 
starts to decrease
and
$\alpha >0$, see  Section 3.5  of  \cite{Zaninetti2020a}.
The   asymptotic radius  is 
\begin{eqnarray}
r(t) = {12}^{ \left( \alpha-5 \right) ^{-1}}{r_{{0}}}^{{\frac {\alpha-3}{
\alpha-5}}} 
\times 
\nonumber \\
\left( -4\,r_{{0}}v_{{0}} \left( \alpha-5 \right)  \left( 
t-t_{{0}} \right) \sqrt {9-3\,\alpha}- \left( \alpha-3 \right) 
 \left( \alpha-5 \right) ^{2} \left( t-t_{{0}} \right) ^{2}{v_{{0}}}^{
2}+12\,{r_{{0}}}^{2} \right) ^{- \left( \alpha-5 \right) ^{-1}}
\quad  ,
\label{traiettoriapotenza}
\end{eqnarray}
and the asymptotic velocity
\begin{eqnarray}
v(t) = 
2\, \left( -4\,r_{{0}}v_{{0}} \left( \alpha-5 \right)  \left( t-t_{{0}
} \right) \sqrt {9-3\,\alpha}- \left( \alpha-3 \right)  \left( \alpha-
5 \right) ^{2} \left( t-t_{{0}} \right) ^{2}{v_{{0}}}^{2}+12\,{r_{{0}}
}^{2} \right) ^{{\frac {4-\alpha}{\alpha-5}}}
\times
\nonumber \\
\left( 2\,{r_{{0}}}^{{\frac {2\,\alpha-8}{\alpha-5}}}\sqrt {9-3\,
\alpha}+v_{{0}}{r_{{0}}}^{{\frac {\alpha-3}{\alpha-5}}} \left( \alpha-
3 \right)  \left( \alpha-5 \right)  \left( t-t_{{0}} \right)  \right) 
{12}^{ \left( \alpha-5 \right) ^{-1}}v_{{0}}
\quad  .
\end{eqnarray}
An example  of trajectory is reported  in Figure 
\ref{trajpowerlaw} 
with  data as in Table \ref{datafitsnr1993j}.

\begin{table}
\caption
{
Numerical values of the parameters
for the fit and  the theoretical models applied to  \snr1993j.
}
 \label{datafitsnr1993j}
 \[
 \begin{array}{ccc}
 \hline
 \hline
 \noalign{\smallskip}
  model& values  & \chi^2        \\
 \noalign{\smallskip}
 \hline
 \noalign{\smallskip}
 Fit~ by~a~ power~law ~  &\alpha_{fit}=0.828;  C=0.015; &  43 \\
 \noalign{\smallskip}
Classic ~ power~law~profile 
&
 \alpha=2.5; r_{0} = 1.0\,10^{-5}~{\mathrm{pc}};
&   
176.6
\\
~ &  t_{0}=5\,10^{-4}~{yr}; v_0=20 000\frac{km}{s}
& ~ 
\\
\noalign{\smallskip} 
Relativistic~NFW   &
 b=0.00185 {\mathrm{pc}}; r_{0} = 1\,10^{-4}~{\mathrm{pc}};
&   
823
\\
~ &  t_{0}=3.6\,10^{-4}~{yr}; v_0= 269813\frac{km}{s}
& ~ 
\\
Relativistic~NCD  &
\delta =1.16; r_{0} = 5\,10^{-5}~{\mathrm{pc}};
&   
9589
\\
~ &  t_{0}=1.8\,10^{-4}~{yr}; v_0= 269813\frac{km}{s}
& ~ 
\\

\noalign{\smallskip} 
\hline\hline
 \end{array}
 \]
 \end {table}

As a consequence, we may derive  an expression 
for the  theoretical   luminosity in presence 
of an inverse  power law profile, $L_{theo}$, based 
on Equations  (\ref{eqnlmtrho})  and (\ref{lcobs}) 
\begin{eqnarray}
L_{theo}
=
\rho_0\,
128\,{r_{{0}}}^{{\frac {-2\,d+5\,\alpha-15}{\alpha-5}}}{v_{{0}}}^{3}{
12}^{{\frac {5-d}{\alpha-5}}}
\times
\nonumber \\
 \left( -4\,r_{{0}}v_{{0}} \left( \alpha-5 \right)  \left( t-t_{{0}}
 \right) \sqrt {9-3\,\alpha}- \left( \alpha-3 \right)  \left( \alpha-5
 \right) ^{2} \left( t-t_{{0}} \right) ^{2}{v_{{0}}}^{2}+12\,{r_{{0}}}
^{2} \right) ^{{\frac {d+10-3\,\alpha}{\alpha-5}}}\pi
\times  
\nonumber \\
\left( r_{{0}}\sqrt {9-3\,\alpha}+{\frac {v_{{0}} \left( \alpha-3
 \right)  \left( \alpha-5 \right)  \left( t-t_{{0}} \right) }{2}}
 \right) ^{3}
\quad .
\label{ltheoclassicpower}
\end{eqnarray}
The above luminosity is based  on theoretical arguments
and no fitting procedure was used.
The observed luminosity, $L_{obs}$, can be obtained introducing 
\begin{equation}
L_{obs} = C_{obs} \times L_{theo}
\label{lobstheo} 
\quad  ,
\end{equation}
where $C_{obs}$ is a constant.
Similarly,
\begin{equation} 
M_{obs} = -\log_{10} (L_{theo}) + k_{obs}  
\quad .
\label{mobstheo}
\end{equation}

\section{The relativistic equation of motion}

\label{section_relativistic}
The  relativistic conservation of kinetic energy in 
the thin layer approximation as derived  in \cite{Zaninetti2020c}
is   
\begin{equation}
M_0(r_0) c^2 (\gamma_0-1)=M(r) c^2 (\gamma-1)  
\quad ,
\label{cons_rel_energy}
\end{equation}
where $M_0(r_0)$ and $M(r)$ are the swept masses at
the two radii $r_0$ and $r$ respectively,
$\gamma_0=\frac{1}{\sqrt{1-\beta_0^2}}$
and   $\beta_0=\frac{v_0}{c}$. 

\subsection{The NFW profile}

\label{motionnfw}
\begin{figure*}
\begin{center}
\includegraphics[width=7cm,angle=-90]{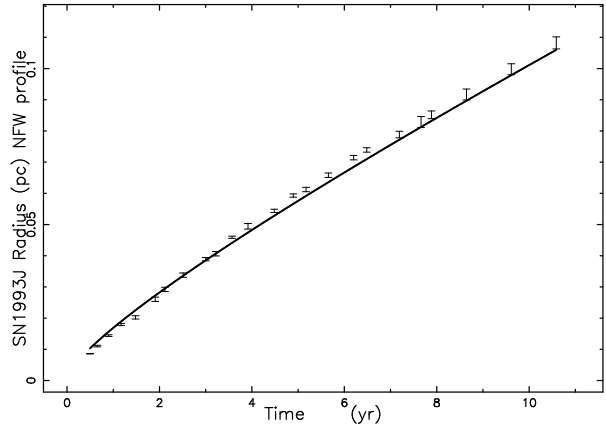}
\end {center}
\caption
{
Numerical  radius for the NFW profile 
(full line),
with data as in 
Table~\ref{datafitsnr1993j}.
The model is the conservation of the relativistic  energy
in the presence of an NFW  profile for the density.
}
\label{nfwrelsn1993j}
    \end{figure*}

\begin{figure*}
\begin{center}
\includegraphics[width=5cm,angle=-90]{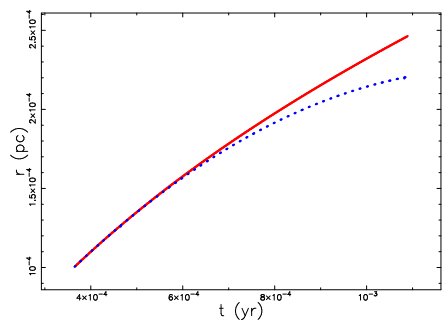}
\end {center}
\caption
{
Numerical solution (full red line) and  
Taylor approximation 
(blue dashed  line)
for the NFW profile with  
parameters  as in
Table~\ref{datafitsnr1993j}.
The model is the conservation of the relativistic  energy
in the presence of an NFW  profile for the density.
}
\label{taylor_theo_nfw}
    \end{figure*}

We  assume that the medium 
around the SN
scales as  the
Navarro--Frenk--White (NFW) profile:
\begin{equation}
 \rho (r;r_0,b)  = \Bigg \{ \begin{array}{ll}
            \rho_c                          & \mbox {if $r \leq r_0 $ } \\
            \frac
             {
              \rho_{{c}}r_{{0}} \left( b+r_{{0}} \right) ^{2}
             }
             {
              r \left( b+r \right) ^{2}
             }
                 & \mbox {if $r >     r_0 $}
            \end{array}
\label{piecewisenfw},
\end{equation}
where 
$\rho_c$ is the density at $r=0$ 
and $r_0$ is the radius after which the density 
starts to decrease,  
see \cite{Navarro1996}.  
The total mass swept, $ M(r;r_0,b,\rho_c) $,
in the interval [0,r]
is
\begin{eqnarray}
M (r;r_0,\rho_c,b)=
{\frac {4\,\rho_{{c}}\pi\,{r_{{0}}}^{3}}{3}}+4\,{\frac {\rho_{{c}}
 \left( b+r_{{0}} \right) ^{2} \left(  \left( b+r \right) \ln  \left( 
b+r \right) +b \right) r_{{0}}\pi}{b+r}}
\nonumber  \\
-4\,\rho_{{c}} \left( b+r_{{0}
} \right)  \left(  \left( b+r_{{0}} \right) \ln  \left( b+r_{{0}}
 \right) +b \right) r_{{0}}\pi
\quad .
\end{eqnarray}
Inserting the above mass in equation (\ref{cons_rel_energy})
makes it  possible  to derive the velocity of the
trajectory as a function of the radius  as well as the
differential equation of the first order which regulates the 
motion.
The differential equation  has a complicated behaviour
which  is not presented and Figure~\ref{nfwrelsn1993j}
displays the numerical solution.
Conversely, we present  an approximate solution
as  a third-order Taylor series expansion
about $r=r_0$  
\begin{eqnarray}
r(t;r_0,v_0,t_0,b) 
=
{\frac {1}{2\,r_{{0}}c} \Bigg  ( 3\,   ( t-t_{{0}}   ) ^{2}
   ( -v_{{0}}+c   ) ^{2}   ( v_{{0}}+c   ) ^{2}\sqrt {
   ( {c}^{2}-{v_{{0}}}^{2}   ) ^{-1}}
}
\nonumber \\
{
-3\,c  \Big (    ( t-t_{
{0}}   ) ^{2}{c}^{2}-{t_{{0}}}^{2}{v_{{0}}}^{2}+   ( 2\,t{v_{{0
}}}^{2}+2/3\,r_{{0}}v_{{0}}   ) t_{{0}}-{t}^{2}{v_{{0}}}^{2}-2/3\,
tr_{{0}}v_{{0}}-2/3\,{r_{{0}}}^{2} \Big  ) \Bigg   ) }
\quad  .
\end{eqnarray}
Figure \ref{taylor_theo_nfw} presents
the Taylor approximation of the trajectory 
in the
restricted range of time  $[4\,10^{-4}\,yr-10^{-3}\,yr]$.

\subsection{NCD case}

\label{sectionncd}

\begin{figure*}
\begin{center}
\includegraphics[width=7cm,angle=-90]{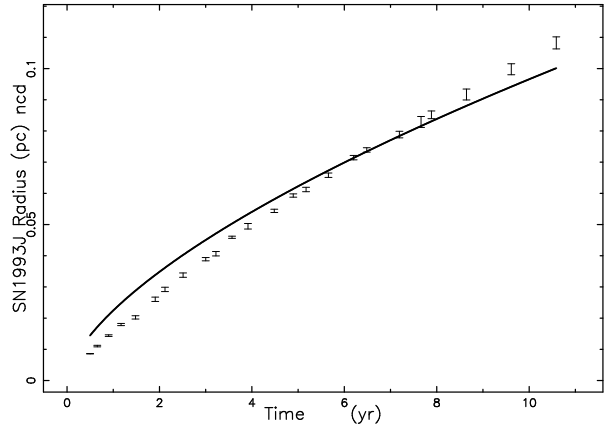}
\end {center}
\caption
{
Numerical  solution  of the differential equation
(\ref{diffeqnncd}) 
 for the NCD case 
(full line),
with data as in 
Table~\ref{datafitsnr1993j}.
The  astronomical data of \snr1993j
are represented with vertical error bars.
The model is the conservation of the relativistic  energy
in the NCD case.
}
\label{ncdrelsn1993j}
    \end{figure*}

\begin{figure*}
\begin{center}
\includegraphics[width=5cm,angle=-90]{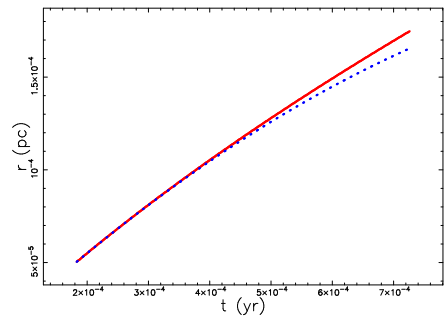}
\end {center}
\caption
{
Numerical solution (full red line) and  
Taylor approximation
(blue dashed  line)
for the NCD  case  with  
parameters  as in
Table~\ref{datafitsnr1993j}.
The model is the conservation of the relativistic  energy
in the NCD case.
}
\label{taylor_theo_ncd}
    \end{figure*}

We assume that the swept  mass  scales as  
\begin{equation}
 M(r;r_0,\delta)  = \Bigg \{ \begin{array}{ll}
M_0                            & \mbox {if $r \leq r_0 $ } \\
M_0 (\frac{r}{r_0})^{\delta}   & \mbox {if $r >     r_0 $}
            \end{array}
\label{piecewisencd},
\end{equation}
where 
$M_0$ is the swept mass at  $r=0$,
$r_0$ is the radius after which the swept mass 
starts to increase 
and $\delta$ is a regulating parameter less than 3.
The  differential equation  of the first order 
which regulates
the motion  is obtained inserting the above $M(r)$ 
in equation (\ref{cons_rel_energy})
\begin{equation}
\frac{dr(t;r_0,v_0,c,\delta)}{dt}=
\frac{AN}{AD}
\quad ,
\label{diffeqnncd}
\end{equation}
where
\begin{eqnarray}
AN=
\Bigg ( {16\,   ( c-v_{{0}}   ) c   ( {r_{{0}}}^{-2\,\delta}
   ( -5/8\,{c}^{2}+5/8\,{v_{{0}}}^{2}   )    ( r   ( t
   )    ) ^{2\,\delta}+{r_{{0}}}^{-3\,\delta}   ( 1/8\,{c}
^{2}-1/8\,{v_{{0}}}^{2}   )    ( r   ( t   )    ) ^{
3\,\delta}
}
\nonumber \\
{
+   ( r   ( t   )    ) ^{\delta}   ( {c}^{
2}-3/4\,{v_{{0}}}^{2}   ) {r_{{0}}}^{-\delta}-1/2\,{c}^{2}+1/4\,{
v_{{0}}}^{2}   )    ( c+v_{{0}}   ) \sqrt {   ( {c}^{2
}-{v_{{0}}}^{2}   ) ^{-1}}
}
\nonumber \\
{
+   ( 10\,{c}^{4}-15\,{c}^{2}{v_{{0
}}}^{2}+5\,{v_{{0}}}^{4}   ) {r_{{0}}}^{-2\,\delta}   ( r
   ( t   )    ) ^{2\,\delta}-2\,{r_{{0}}}^{-3\,\delta}
   ( c-v_{{0}}   ) ^{2}   ( c+v_{{0}}   ) ^{2}   ( 
r   ( t   )    ) ^{3\,\delta}
}
\nonumber \\
{
+   ( -16\,{c}^{4}+20\,{c
}^{2}{v_{{0}}}^{2}-4\,{v_{{0}}}^{4}   )    ( r   ( t
   )    ) ^{\delta}{r_{{0}}}^{-\delta}+8\,{c}^{4}-8\,{c}^{2}
{v_{{0}}}^{2}+{v_{{0}}}^{4}} \Bigg)^{1/2} c
\quad ,
\end{eqnarray}
and
\begin{eqnarray}
AD=
2\,c   ( c-v_{{0}}   )    ( c+v_{{0}}   )    ( {r_
{{0}}}^{-\delta}   ( r   ( t   )    ) ^{\delta}-1
   ) \sqrt {   ( {c}^{2}-{v_{{0}}}^{2}   ) ^{-1}}
\nonumber \\
+{r_{{0}}
}^{-2\,\delta}   ( {c}^{2}-{v_{{0}}}^{2}   )    ( r   ( 
t   )    ) ^{2\,\delta}+   ( -2\,{c}^{2}+2\,{v_{{0}}}^{2}
   )    ( r   ( t   )    ) ^{\delta}{r_{{0}}}^{-
\delta}+2\,{c}^{2}-{v_{{0}}}^{2}
\quad  .
\end{eqnarray}
The above differential does not have an analytical solution
and therefore the solution should be derived numerically
except 
about $r=r_0$ where a third-order Taylor series expansion gives 
\begin{eqnarray}
r(t;r_0,v_0,t_0,\delta)=
r_{{0}}+v_{{0}} \left( t-t_{{0}} \right) 
\nonumber  \\
+{\frac {\delta\, \left( c-v
_{{0}} \right)  \left( c+v_{{0}} \right)  \left( t-t_{{0}} \right) ^{
2}}{2\,cr_{{0}}} \left( {c}^{2}-c\sqrt {{c}^{2}-{v_{{0}}}^{2}}-{v_
{{0}}}^{2} \right) {\frac {1}{\sqrt {{c}^{2}-{v_{{0}}}^{2}}}}}
\quad .
\label{rt_taylor_ncd}
\end{eqnarray}
Figure \ref{ncdrelsn1993j} 
presents the numerical solution
and Figure \ref{taylor_theo_ncd} 
the Taylor approximation of the trajectory.

\section{Astrophysical results}

\label{section_astrophysical}

We introduce  one  SN and two GRBs which 
were  processed.

\subsection{The case of \snr1993j}

\begin{figure}
\begin{center}
\includegraphics[width=6cm,angle=-90]{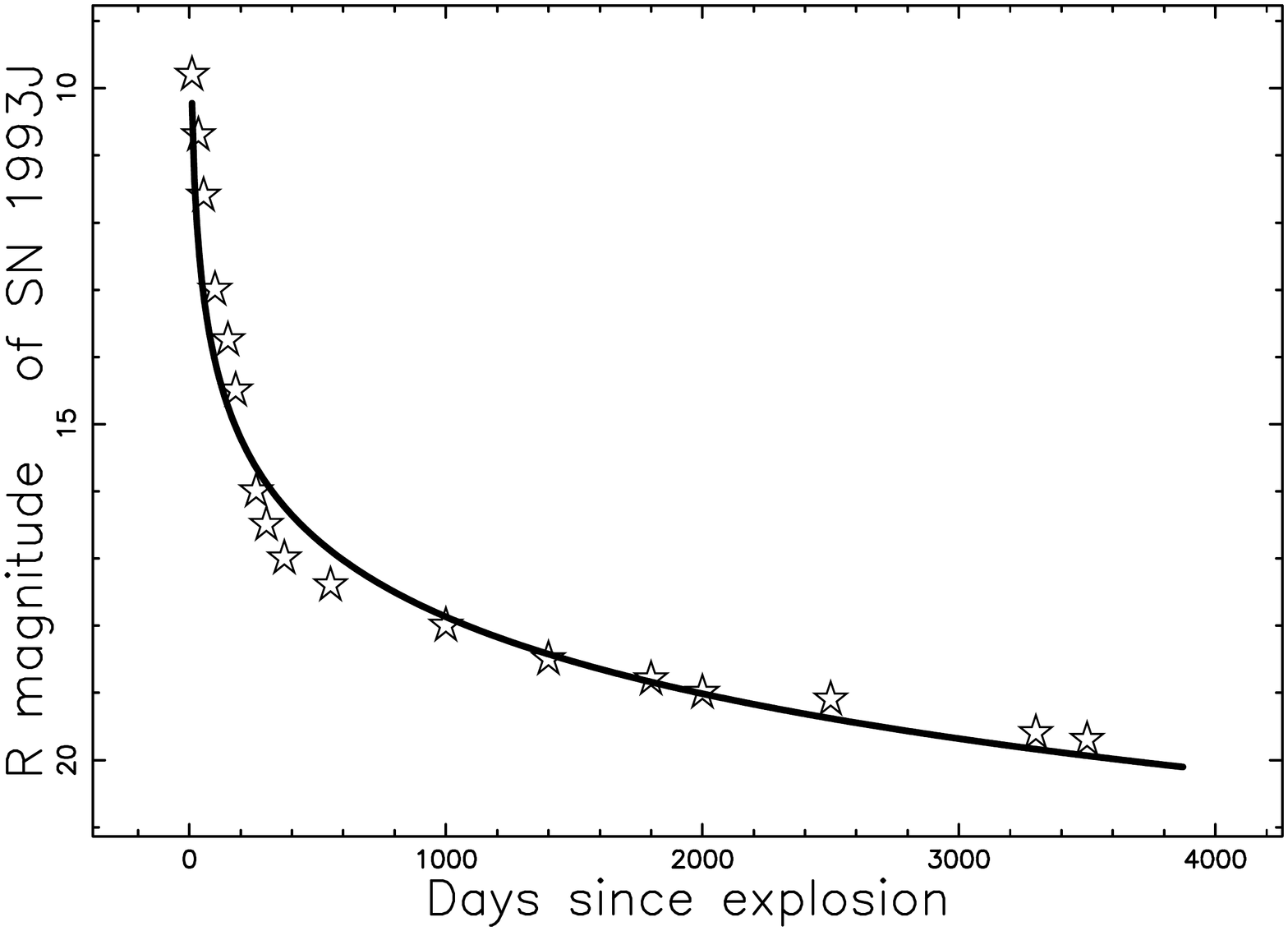}
\end {center}
\caption
{
The $R$ LC of \snr1993j over 10 yr (empty stars)
and theoretical curve 
in the classical framework of a power law 
profile  for the density 
as given by eq.  
(\ref{lobstheo}) 
(full
line). 
Parameters of the trajectory as in Table \ref{datafitsnr1993j}, 
$d$ = 6, $k_{obs}$ = -11.5 and
$\rho_0$ = 1. The data were extracted by the author from Figure 5
in Zhang et al. (2004).
} \label{mtpowerlaw}
    \end{figure}

In this  subsection 
we adopt  a  classical equation of motion with a power law profile of density,
see Section  \ref{motionpowerlaw}.
Figure \ref{mtpowerlaw}
presents the decay of the $R$   magnitude of
\snr1993j, which is type IIb, as well our theoretical curve.

\begin{figure}
\begin{center}
\includegraphics[width=6cm,angle=-90]{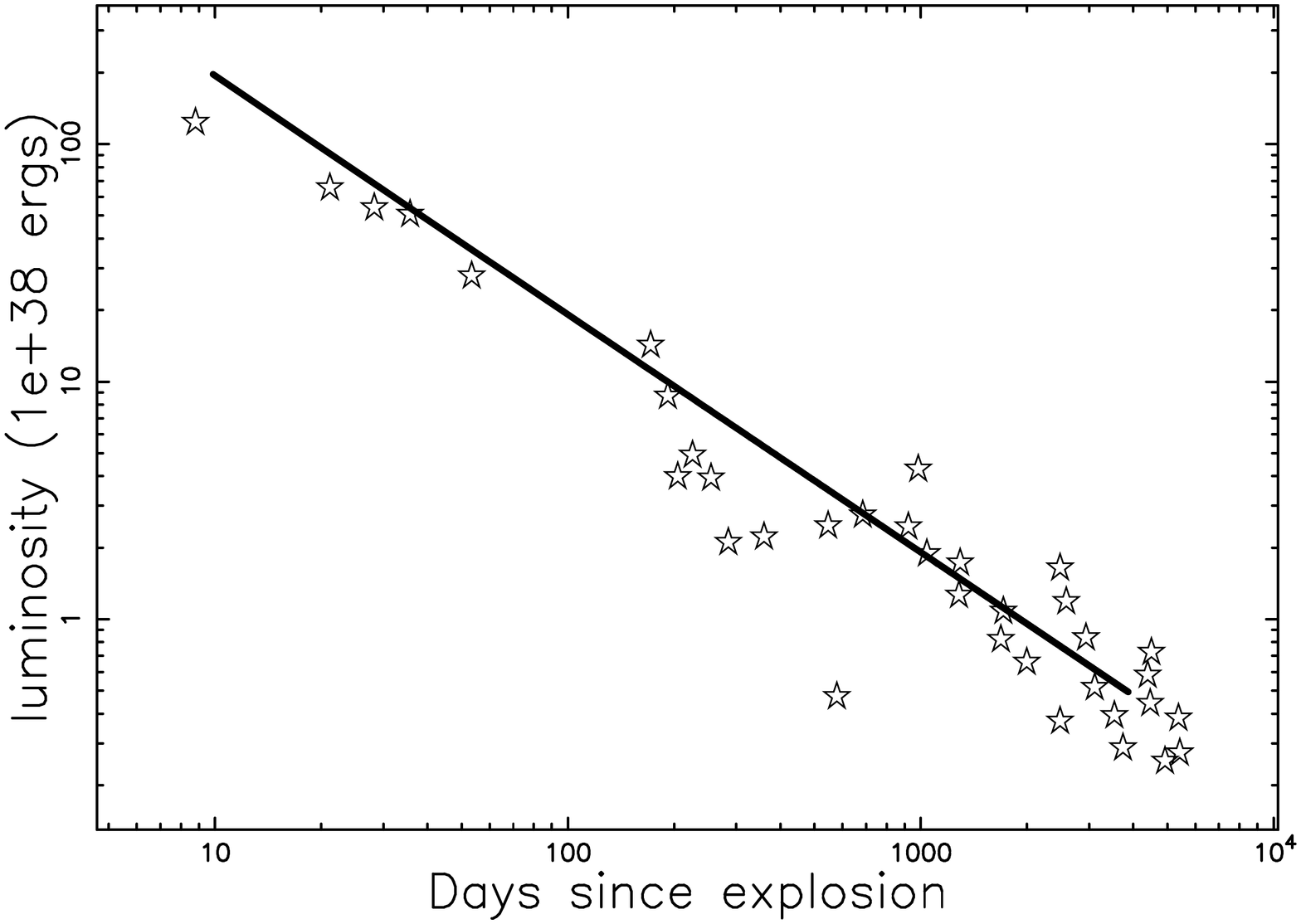}
\end {center}
\caption
{
The  $H-\alpha$ 
and the  2.0--8.0 keV luminosities  of \snr1993j over 10 yr (empty stars)
and theoretical curve 
in the classical framework of a power law 
profile  for the density 
as given by eq.  
(\ref{mobstheo}) 
(full
line). 
Parameters of the trajectory as in Table \ref{datafitsnr1993j}, 
$d$ = 2.5, $C_{obs} = 1.5\,10^{17}$ and
$\rho_0$ = 1. The data were extracted by the author from Figure 5
in \cite{Chandra2009}.
} \label{ltpowerlaw}
    \end{figure}

\begin{figure}
\begin{center}
\includegraphics[width=6cm,angle=-90]{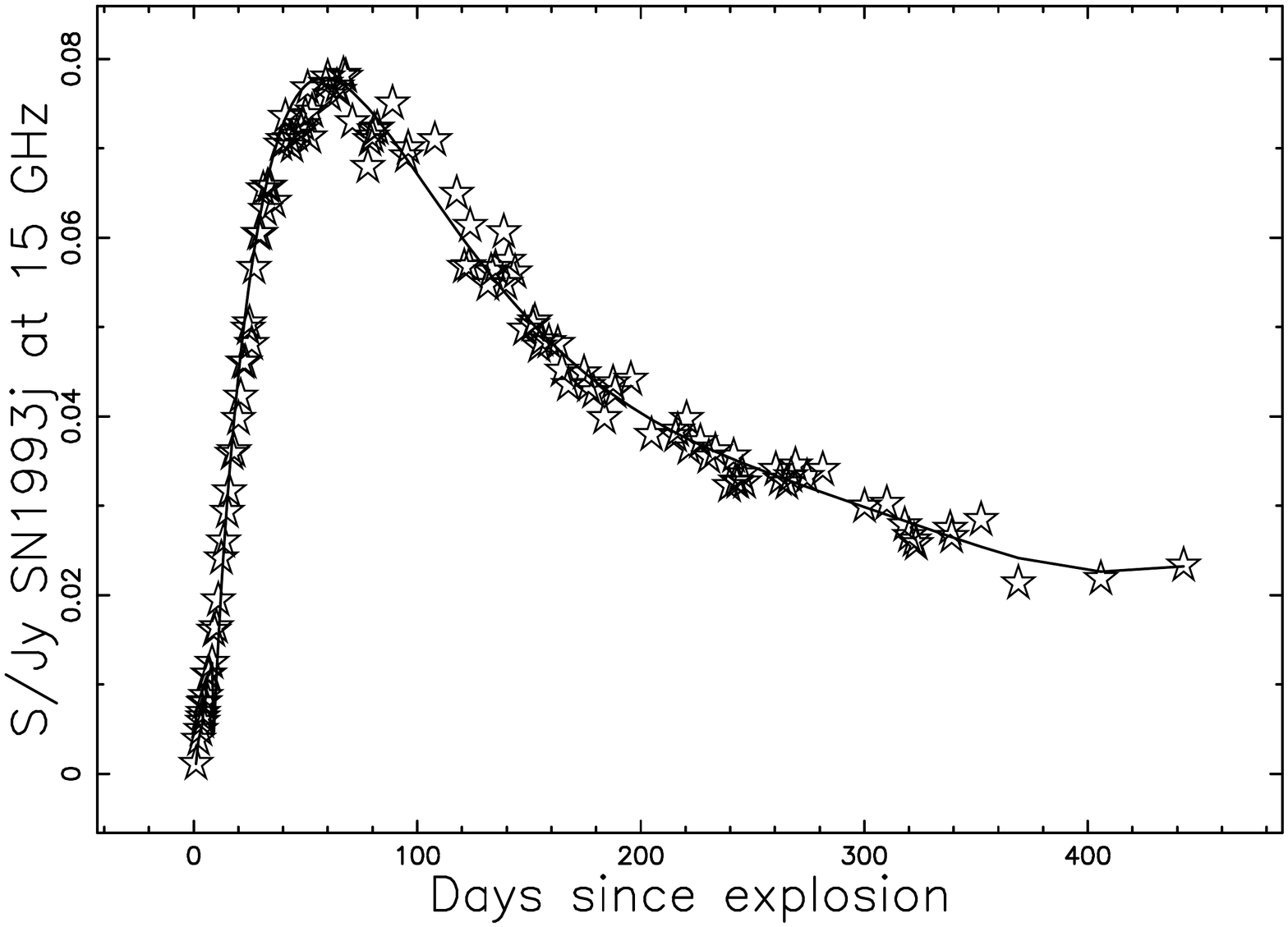}
\end {center}
\caption
{
The radio flux density
of \snr1993j over 443 days (empty stars)
and theoretical behaviour  
in the classical framework of a power law 
profile  for the density 
evaluated with formula (\ref{lcobstau}) (full line).
Parameters of the trajectory as in Table \ref{datafitsnr1993j}
, $d$ = 2.5, $C_{obs} = 8.45\,10^{14}$ and
$\rho_0$ = 1.
} 
\label{radiocomptaunu}
    \end{figure}

\begin{figure}
\begin{center}
\includegraphics[width=6cm,angle=-90]{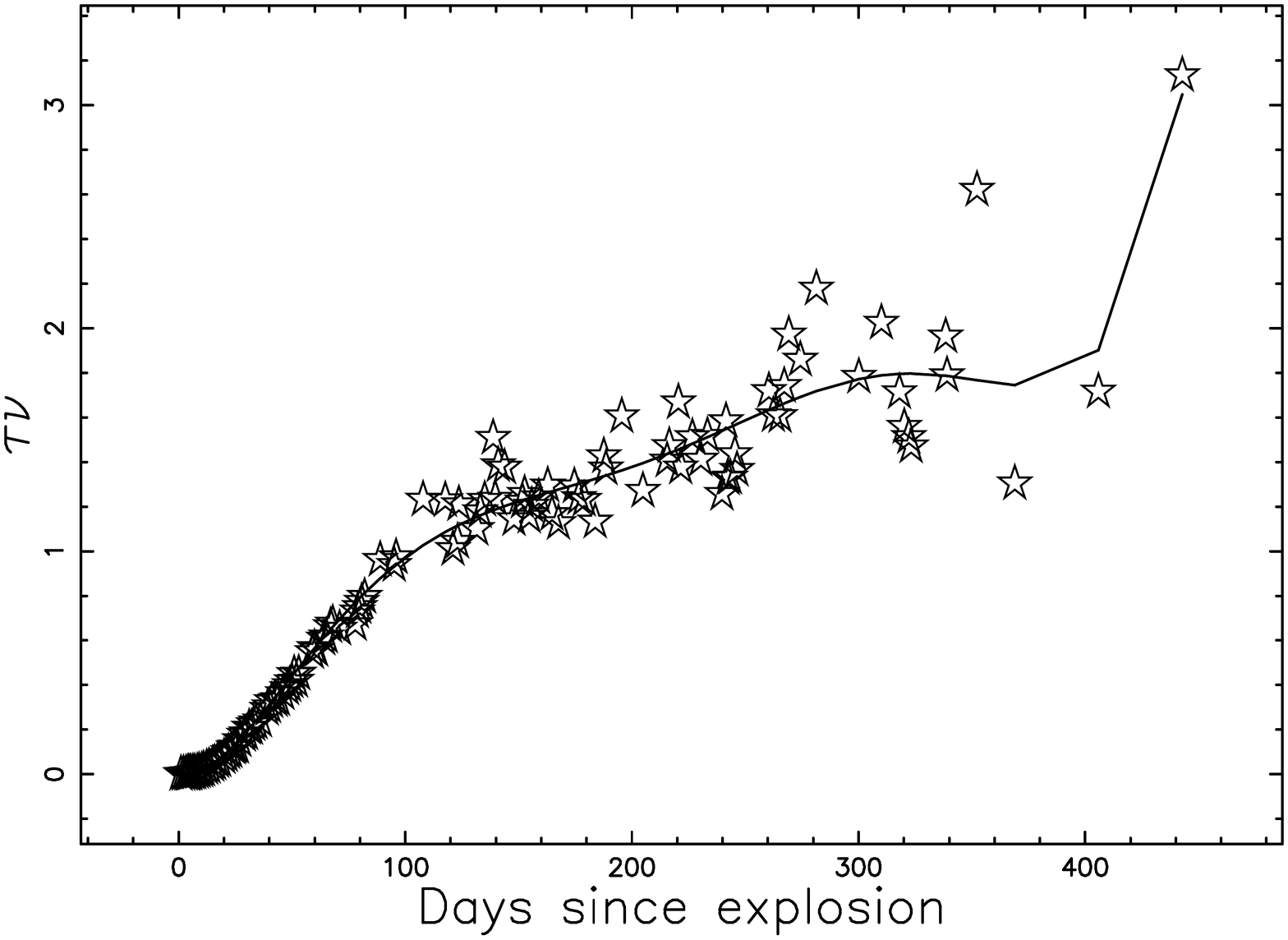}
\end {center}
\caption
{
The time dependence of $\tau_{\nu}$ (empty stars) 
and   
a 
polynomial approximation of  degree  6  (full line).
Parameters as in Figure \ref{radiocomptaunu}.
} 
\label{taunuradio}
    \end{figure}

We present the
$H-\alpha$  with soft and hard band X-ray luminosities
as well the theoretical luminosity  
in Figure \ref{ltpowerlaw}.
Figure \ref{radiocomptaunu} presents the radio flux density
of \snr1993j at 15.2 GHz observed by the Ryle Telescope
as well the theoretical flux,
which requires a time dependent  evaluation 
of the optical depth $\tau_{\nu}$,
see Figure \ref{taunuradio}.

\begin{figure}
\begin{center}
\includegraphics[width=6cm,angle=-90]{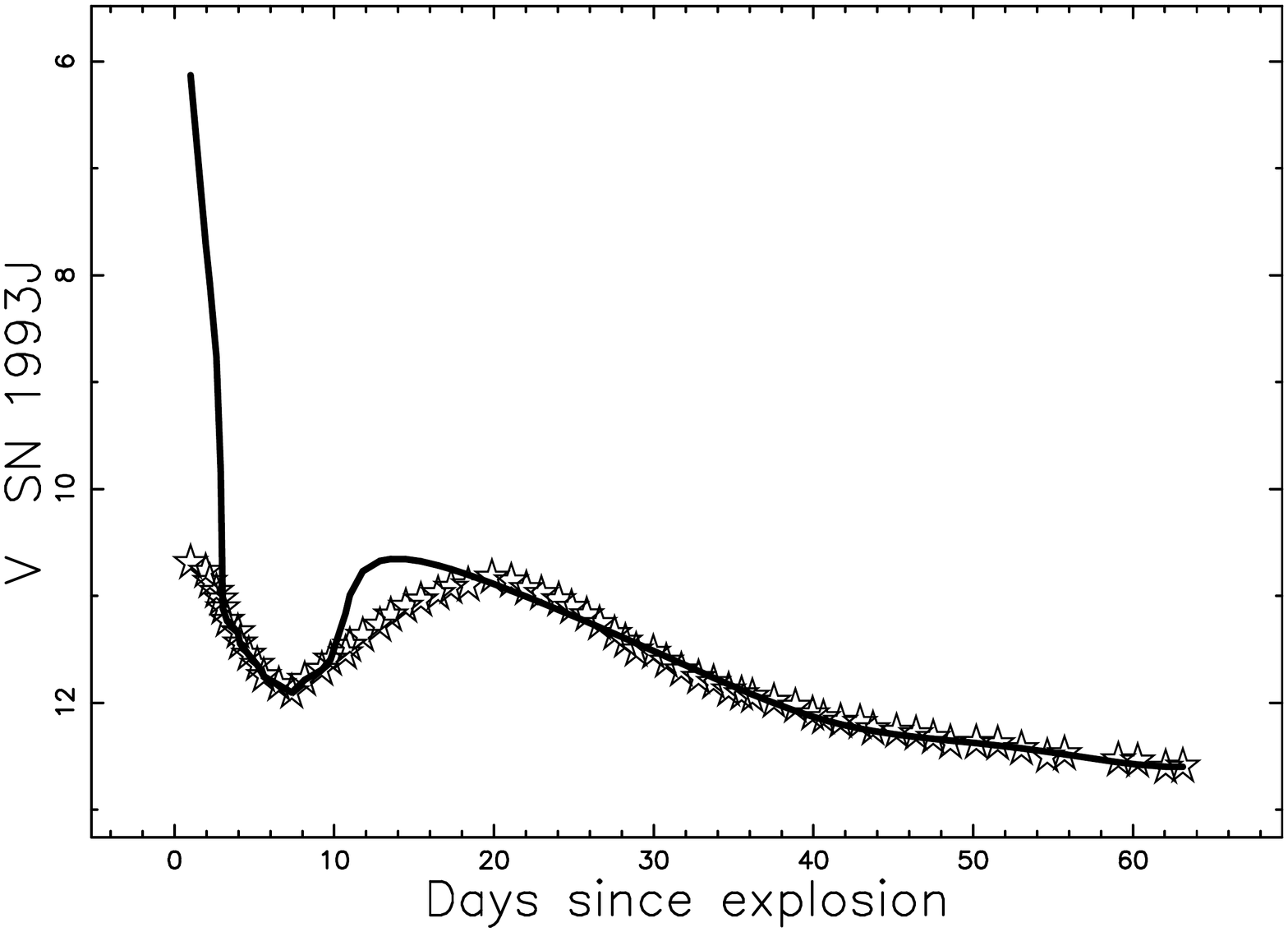}
\end {center}
\caption
{
The $V$ LC of \snr1993j over 63 days (empty stars)
and theoretical curve 
in the classical framework of a power law 
profile  for the density 
as given by eq. (\ref{mobstau}) 
(full
line). 
Parameters of the trajectory as in Table \ref{datafitsnr1993j}, 
$d$ = 6, $k_{obs}$ = -12.5 and
$\rho_0$ = 1. The data were extracted by the author from Figure 4
in \cite{Benson1994}.
} \label{duemagfew}
    \end{figure}

\begin{figure}
\begin{center}
\includegraphics[width=6cm,angle=-90]{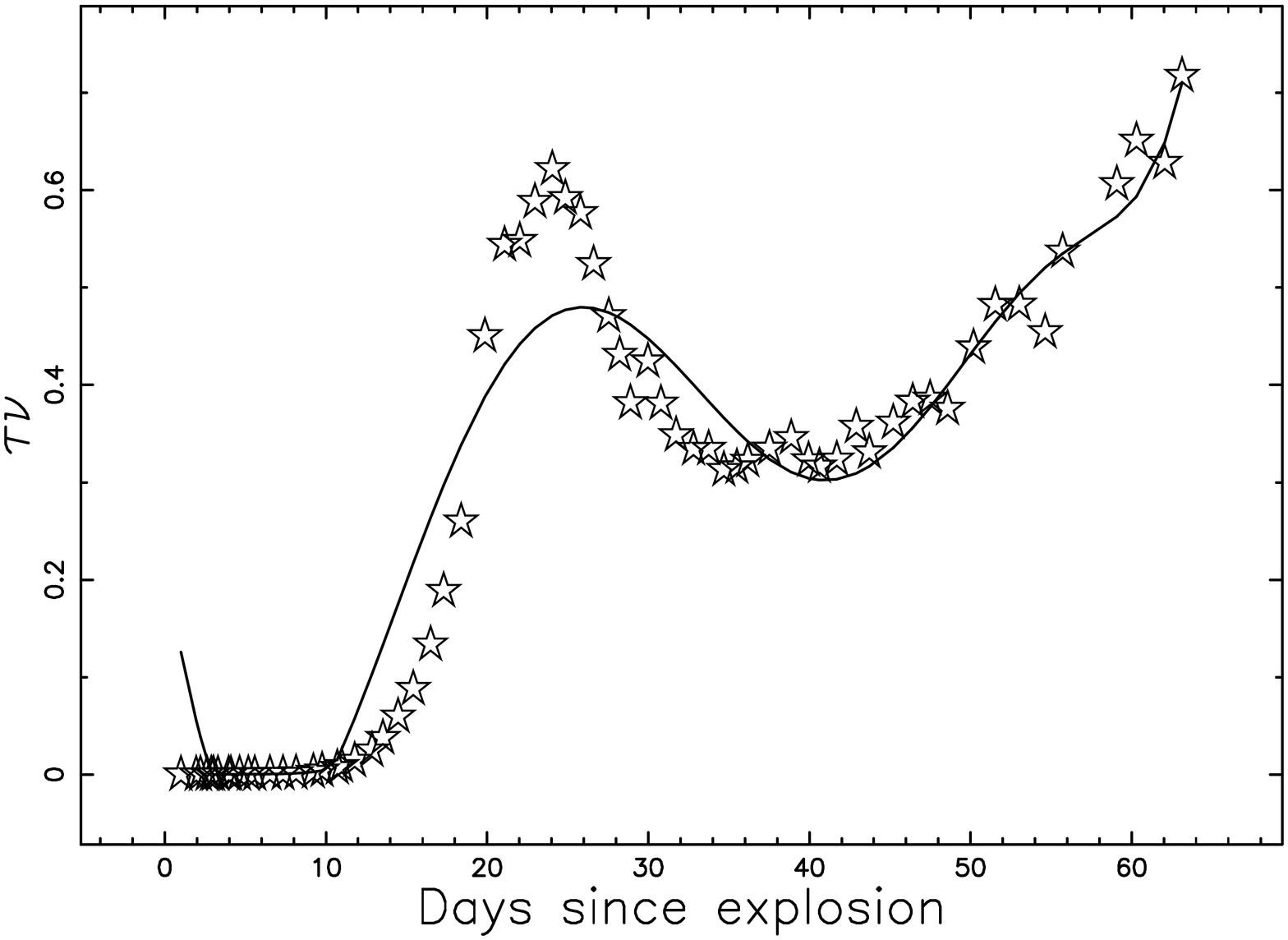}
\end {center}
\caption
{
The time dependence of $\tau_{\nu}$ (empty stars) 
and   
a polynomial approximation
of  degree  10    (full line).
Parameters as in Figure \ref{duemagfew}.
} 
\label{taunufewv}
    \end{figure}

Figure \ref{duemagfew} presents the V-magnitude 
of \snr1993j for few days 
as well the theoretical magnitude
and the time evolution  
of the optical depth $\tau_{\nu}$,
see Figure \ref{taunufewv}.

\subsection{The case of \grb}

\begin{figure}
\begin{center}
\includegraphics[width=6cm,angle=-90]{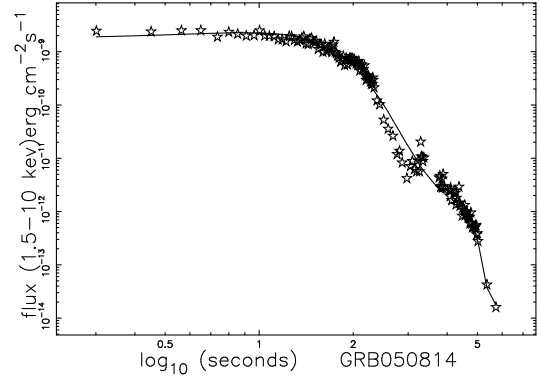}
\end {center}
\caption
{
The XRT flux  of \grb
at 0.2--10 keV  (empty stars)  and
theoretical curve with velocity and radius as given by the 
NFW relativistic numerical model.
The theoretical  luminosity, which is corrected for absorption,
is given by eq. (\ref{lumrel}) (full line).
}
\label{obs_theo_grb}
    \end{figure}

\begin{figure}
\begin{center}
\includegraphics[width=6cm,angle=-90]{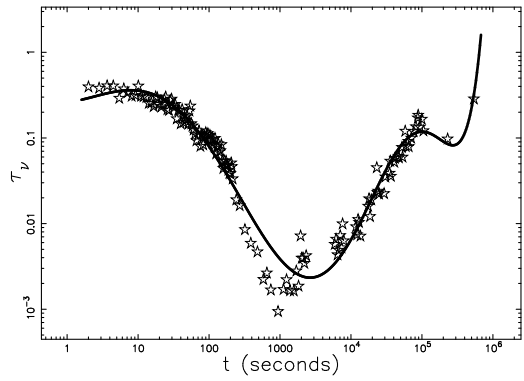}
\end {center}
\caption
{
The time dependence of $\tau_{\nu}$ (empty stars) 
for \grb 
and   
a logarithmic polynomial approximation
of  degree  7  (full line).
Parameters as in Figure \ref{obs_theo_grb}.
} 
\label{tau_grb_fit}
    \end{figure}

In this  subsection 
we adopt  a  relativistic  equation of motion with an NFW profile for the density,
see Section \ref{motionnfw}.
Figure \ref{obs_theo_grb} presents the XRT flux of \grb 
and Figure \ref{tau_grb_fit} presents the 
temporal behaviour  of the optical depth.
 
\subsection{The case of \rgb2}

\begin{figure}
\begin{center}
\includegraphics[width=6cm,angle=-90]{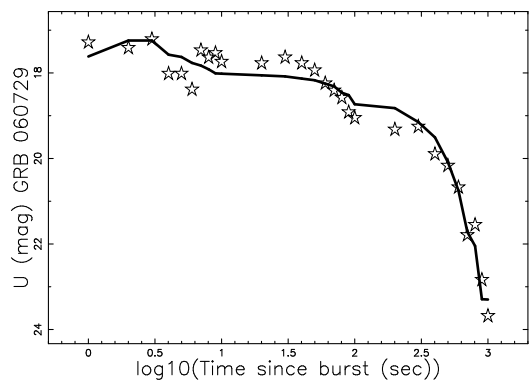}
\end {center}
\caption
{
The LC   of UVOT (U)
+ HST (F330W) for \rgb2
 (empty stars)  and
theoretical curve with radius as given by the 
NCD  relativistic numerical model
with  data as in Table \ref{datafitsnr1993j}.
The theoretical  luminosity is given by eq. (\ref{lumrel})
(full line).
}
\label{ncd_theo_obs_rgb}
    \end{figure}

\begin{figure}
\begin{center}
\includegraphics[width=6cm,angle=-90]{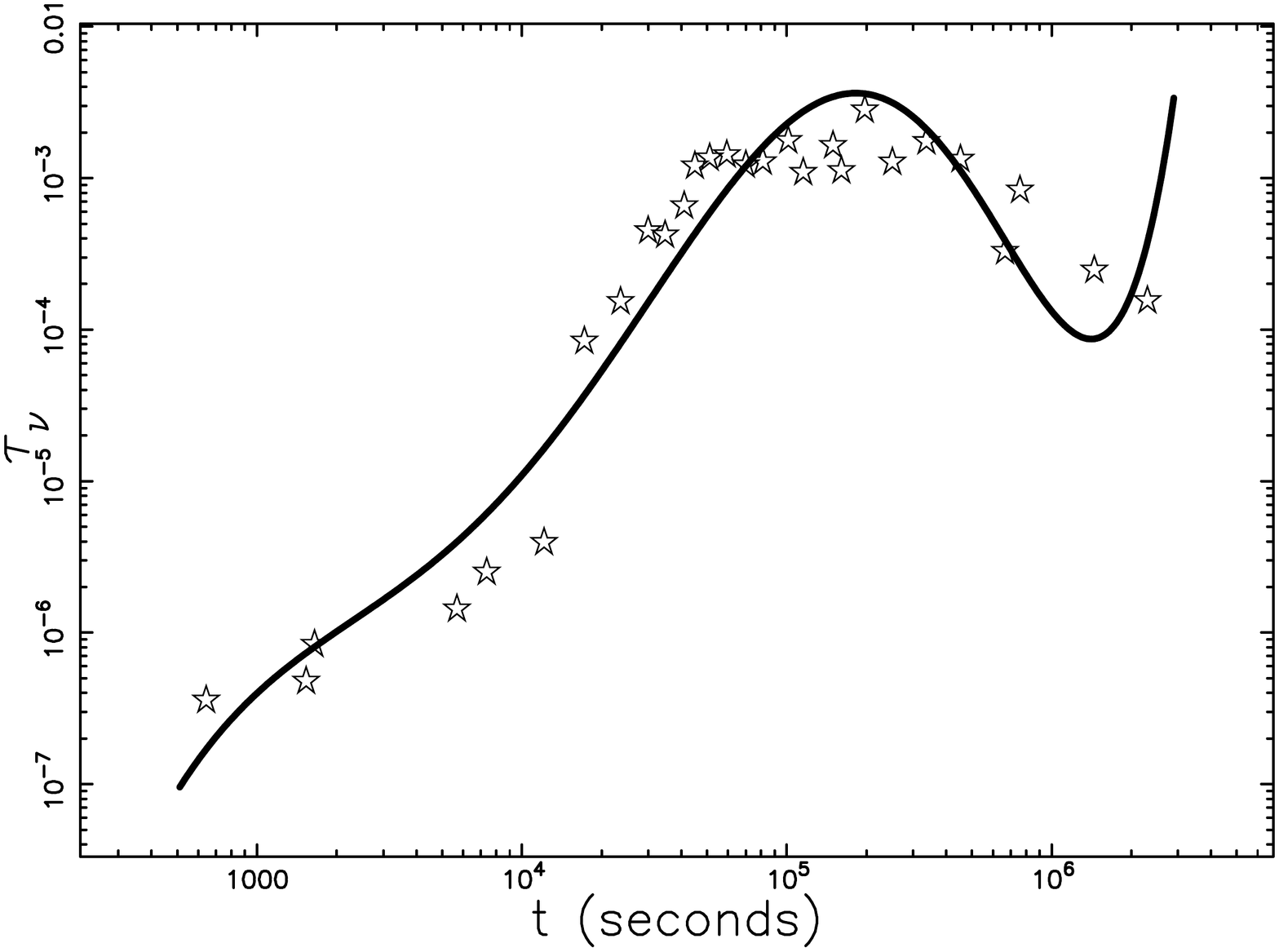}
\end {center}
\caption
{
The time dependence of $\tau_{\nu}$ (empty stars) 
for \rgb2
and   
a logarithmic polynomial approximation
of  degree  10 (full line).
Parameters as in Figure \ref{obs_theo_grb}.
} 
\label{ncd_tau_rgb_fit}
    \end{figure}

In this  subsection 
we adopt  a  relativistic  equation of motion for  the NCD case,
see Section \ref{sectionncd}.
Figure \ref{ncd_theo_obs_rgb} presents 
the LC  of UVOT (U) apparent magnitude
for \rgb2 
and Figure \ref{ncd_tau_rgb_fit} presents the 
temporal behaviour of the optical depth.

\section{Acceleration and magnetic field}

\label{section_magnetic}

\begin{figure}
\begin{center}
\includegraphics[width=6cm]{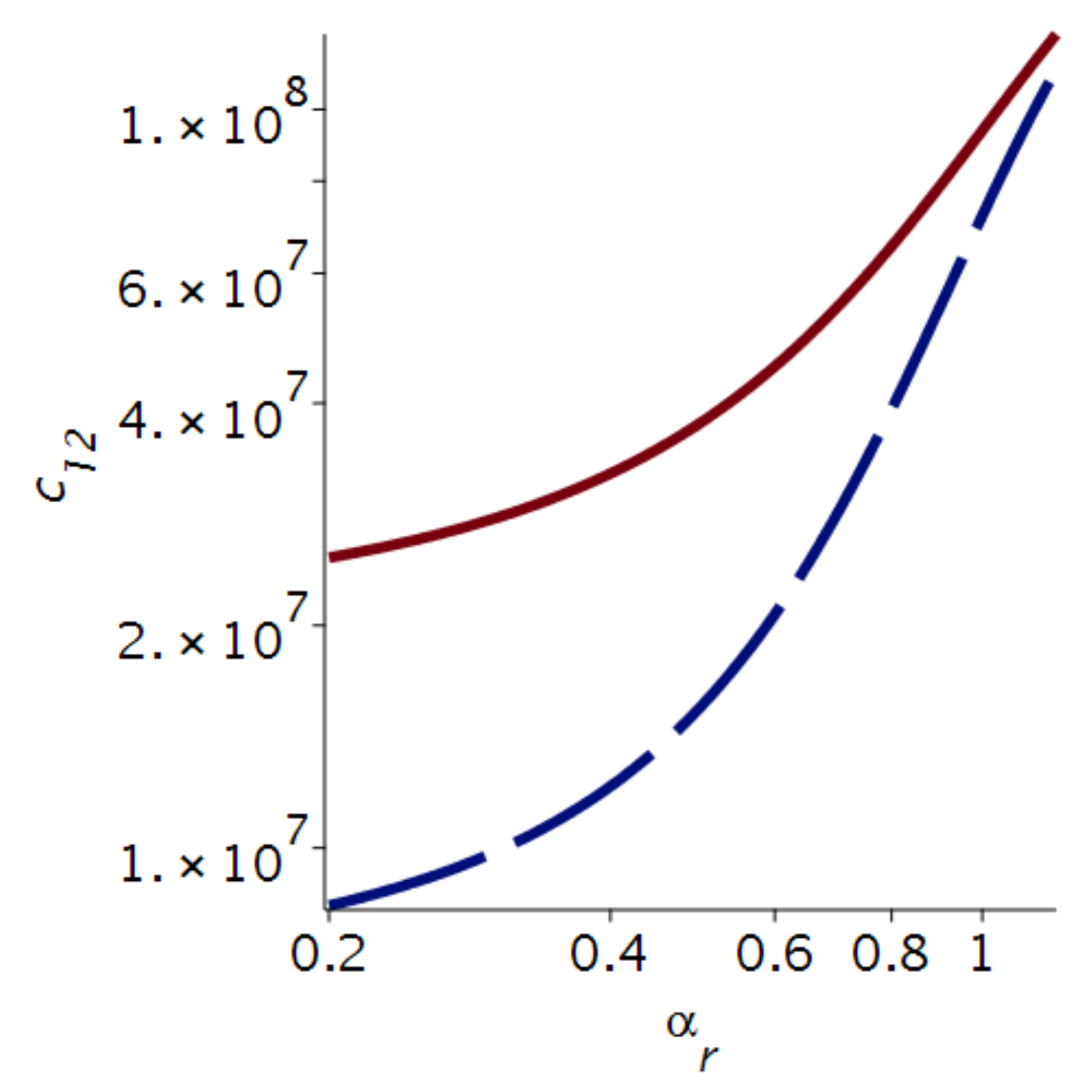}
\end {center}
\caption
{
The constant $c_{12}$
as a function of the spectral index $\alpha_r$ 
when $\nu_1=10^{7} Hz$,  
$\nu_2=10^{10} Hz$        (red  full line)
and  $\nu_2=10^{11} Hz$   (blue dashed  line).
} 
\label{tablec12}
    \end{figure}

\begin{figure}
\begin{center}
\includegraphics[width=6cm,angle=-90]{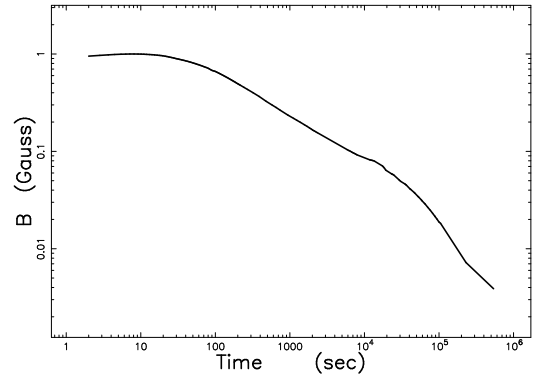}
\end {center}
\caption
{
The time dependence (seconds) of 
the minimum magnetic field 
for \grb 
with theoretical luminosity  as in Figure \ref{tau_grb_fit}
when  $H^{min}=1\,gauss$ at $t=t_0$.
The model is a  fit to a  power law.
} 
\label{magnetic_grb}
    \end{figure}

The flux  at frequency, $S_{\nu}$, in the radio band  for SNs 
is  parametrized
by  
\begin{equation}
S_{\nu} = C_{\nu} \nu^{-\alpha_r}
\quad ,
\end{equation}
where $\alpha_r$ is the observed spectral index
and $C_{\nu}$ is a constant.
As a consequence, the 
luminosity, $L_{\nu}$,
is 
\begin{equation}
L_{\nu} = 4 \pi D^2 S_{\nu}
\quad ,
\end{equation}
where $D$ is the distance.
We now explain how it is possible to derive
the magnetic field
from the luminosity. 
The magnetic field for which the total energy of a radio source
has a  minimum  is
\begin{equation}
H^{min} =
 1.5368\,{\frac {{c_{{12}}}^{2/7}{L}^{2/7} \left(  1.0+k \right) 
^{2/7}}{{\Phi}^{2/7}{R}^{6/7}}}
\quad  
gauss 
\quad  ,
\label{bminpach}
\end{equation}
where 
\begin{equation}
c_{12}=
2\,{\frac {\sqrt {c_{{1}}} \left( -1+\alpha_r \right)  \left( \sqrt {\nu
_{{1}}}{\nu_{{2}}}^{\alpha_r}-{\nu_{{1}}}^{\alpha_r}\sqrt {\nu_{{2}}}
 \right) }{c_{{2}} \left( \nu_{{1}}{\nu_{{2}}}^{\alpha_r}-\nu_{{2}}{\nu_
{{1}}}^{\alpha_r} \right)  \left( -1+2\,\alpha_r \right) }}
\quad ,
\end{equation}
where $\alpha_r$ is the spectral index,
$c_{1}$ and $c_{2}$   are two  constants,
$\nu_{1}$ and $\nu_2$ are the lower and upper frequency 
of synchrotron emission,
$L$ is the luminosity in ${erg}\,{s^{-1}}$,
$k$ is the ratio between energy in heavy particle and electron
energy,
$\Phi$ is  the fraction of source's volume occupied by 
the relativistic electrons and the magnetic field,
and $R$ is the radius of the source; for more details  
see formula (7.14) in \cite{Pacholczyk1970}
or  formula (5.109) in \cite{Condon2016}.
The constant $c_{12}$ is numerically evaluated 
in Table 8 of  \cite{Pacholczyk1970}
and an example is presented in Figure \ref{tablec12}.
The scaling of the magnetic of equipartition
as given by equation (\ref{bminpach}) is
\begin{equation}
H^{min} \propto 
\frac
{{L}^{2/7}}
{{R}^{6/7}}
\quad . 
\label{bscaling}
\end{equation}
The first example presents the temporal behaviour of  $H^{min}$
for  \grb in which we inserted the theoretical
luminosity corrected for absorption, see Figure \ref{magnetic_grb}.

\begin{figure}
\begin{center}
\includegraphics[width=6cm]{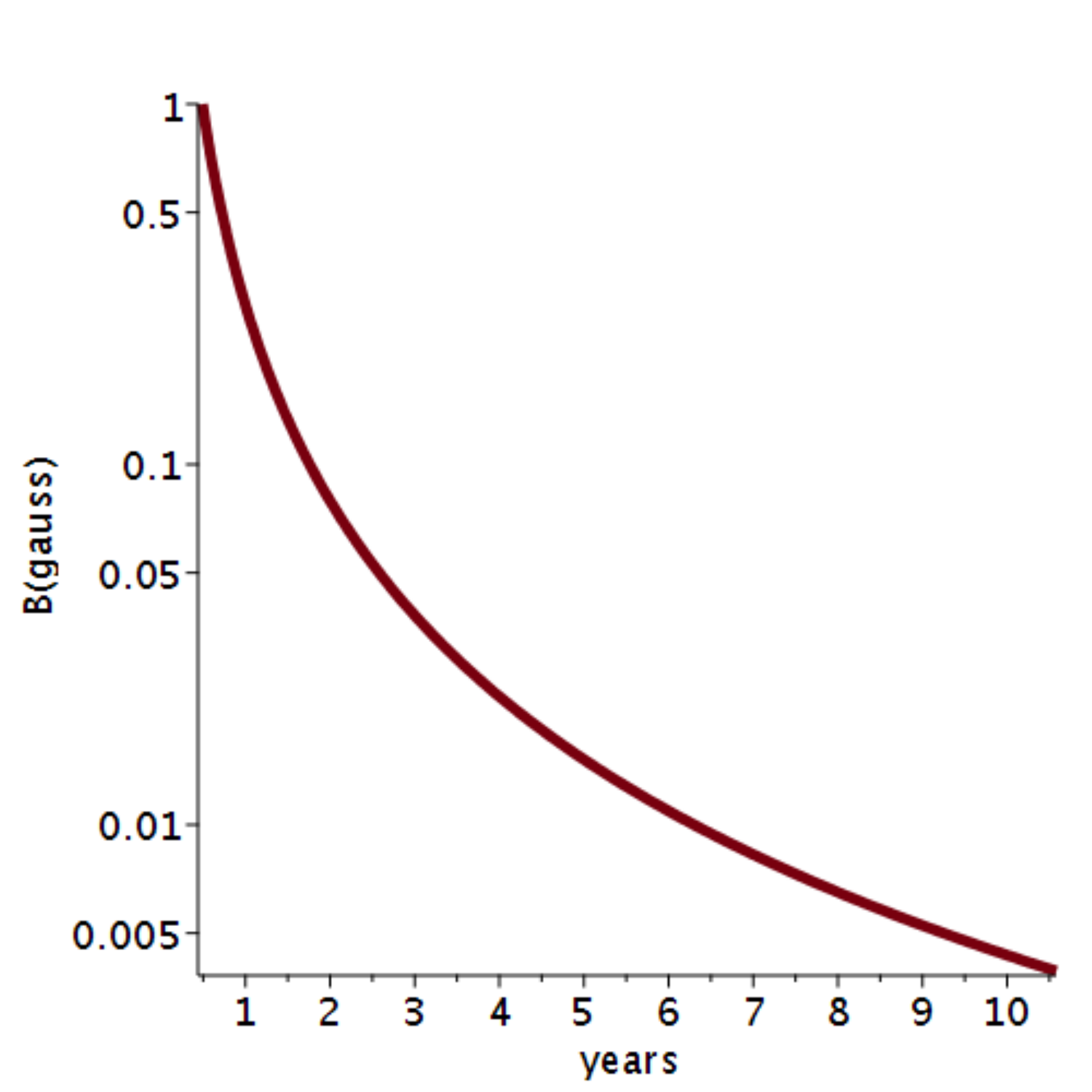}
\end {center}
\caption
{
The time dependence (years) of 
the minimum magnetic field 
for \snr1993j 
with theoretical luminosity  as 
given by   formula (\ref{eqnlmtrho}) 
when  $H^{min}=1\,gauss$ at $t=t_0$.
The model is a  fit to a  power law.
} 
\label{magneticsn1993j}
    \end{figure}

The second example presents the temporal behaviour of  $H^{min}$
for  \snr1993j  in which we inserted the theoretical
luminosity  as given by the power law fit,
see Figure \ref{magneticsn1993j}.

An electron which  loses
its  energy  due to the 
synchrotron radiation
has a lifetime  of
\begin{equation}
\tau_r  \approx  \frac{E}{P_r} \approx  500  E^{-1} H^{-2} sec
\quad ,
\label {taur}
\end{equation}
where
$E$  is the energy in ergs,
$H$ the magnetic field in Gauss,
and
$P_r$  is the total radiated
power, see Eq. 1.157 in  \cite{lang}.
The  energy  is connected  to  the critical
frequency, see Eq. 1.154 in  \cite{lang},
by
\begin {equation}
\nu_c = 6.266 \times 10^{18} H E^2~\mathrm{Hz}
\quad  .
\label {nucritical}
\end{equation}

\begin{figure}
\begin{center}
\includegraphics[width=6cm]{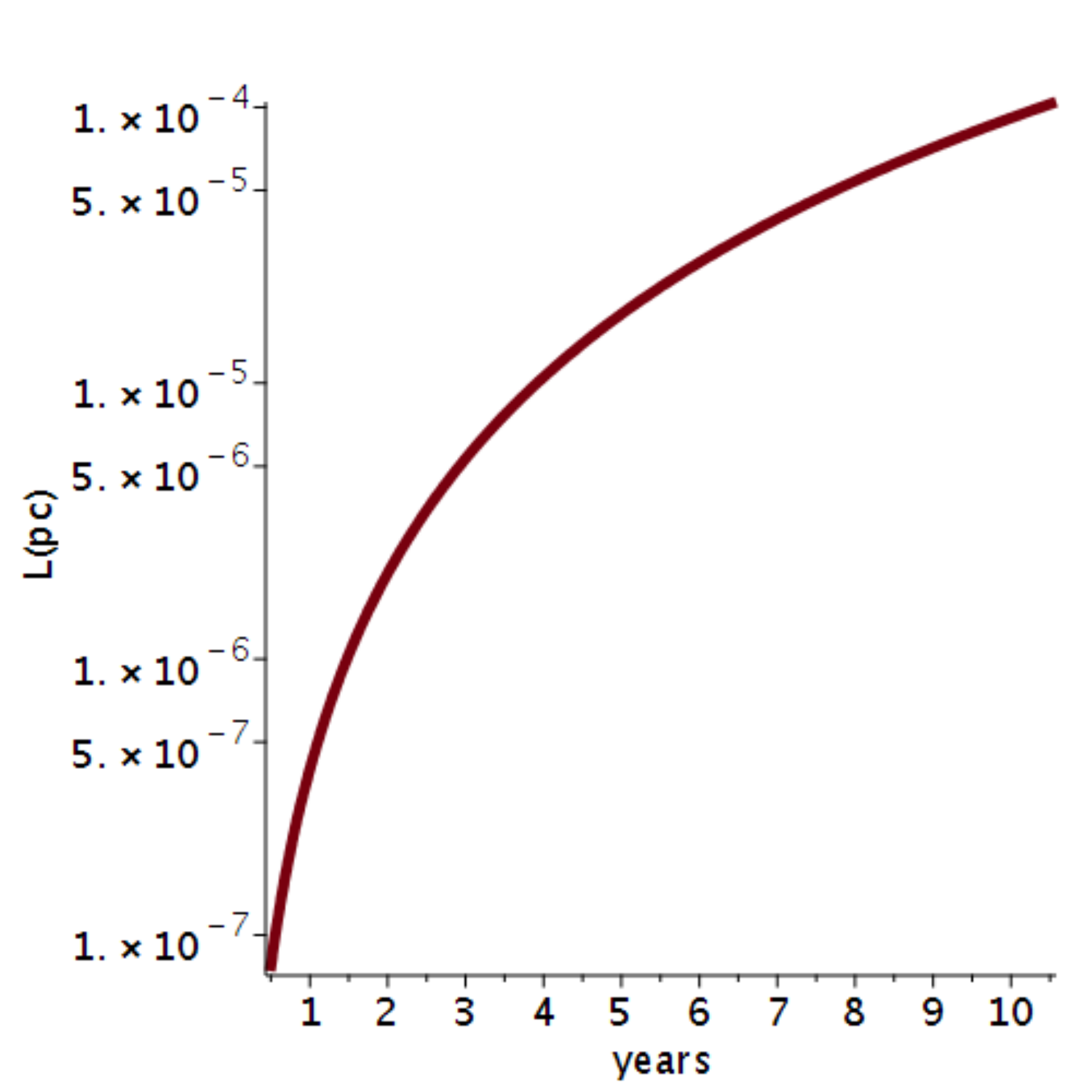}
\end {center}
\caption
{
The time dependence (years) of 
the mean free path length 
for \snr1993j 
when $E=1$\ keV.
The model is a  fit to a  power law.
} 
\label{linequality}
    \end{figure}

The lifetime
for synchrotron  losses is
\begin{equation}
\tau_{syn} =
 39660\,{\frac {1}{H\sqrt {H\nu}}} \, \mathrm{yr}
\quad  .
\end{equation}
Following \cite{Fermi49,Fermi54},
the gain  in  energy  in a continuous    form
is  proportional to its
energy, $E$,
\begin  {equation}
\frac {d  E}  {dt }
=
\frac {E }  {\tau_{II} }   \quad,
\end {equation}
where $\tau_{II}$ is the  typical time scale,
\begin {equation}
\frac{1}{\tau_{II}}  = \frac {4} {3 }
( \frac {u^2} {c^2 }) (\frac {c } {L_{II} })
\quad ,
\label {tau2}
\end   {equation}
where $u$ is the velocity of the accelerating cloud
belonging  to the advancing shell of the GRB,
$c$    is the speed of light 
and $L_{II}$  is the
mean free path between clouds, 
see Eq. 4.439 in \cite{lang}. 
The mean free path between the accelerating clouds
in the Fermi II mechanism can be found from the following
inequality in time:
\begin{equation}
\tau_{II} < \tau_{sync}
\quad  ,
\end{equation}
which  corresponds to  the following  inequality for the
mean free  path  between scatterers
\begin{equation}
L  <
16182.11\,{\frac {{u}^{2}}{{H}^{3/2}\sqrt {\nu}{c}^{2}}}
\,\mathrm{pc}
\quad  .
\end{equation}
The mean free path length for a GRB
which emits synchrotron emission  
around   1 keV   ( $2.417\,10^{17}$ Hz ) 
is 
\begin{equation}
L <
3.2908\,10^{-5}{\frac {{\beta}^{2}}{{H}^{3/2}\sqrt {{\it E(kev)}}}}
\quad pc
\end{equation}
where  
$\beta$ is the velocity 
of the cloud divided  
by the speed of light. 
When this inequality  is fulfilled, the direct
conversion of the rate   of  kinetic energy
into radiation can be adopted.
Figure \ref{linequality} presents 
the above line in the framework of the fitted 
model.
The mean free path length
varies from $1.5\,10^{-5}$ to $1.5\,10^{-4}$
with respect to the numerical value of the advancing radius.

\section{Conclusions}

{\bf Classical and relativistic flux of energy:}
\newline
The classical flux of kinetic energy  
has an analytical expression
in the case of energy conservation 
in   the presence of a power law 
profile for the density, see equation (\ref{ltheoclassicpower}).
The relativistic flux of energy  in the two cases 
here analysed can only be found numerically.
\newline
{\bf Momentum versus energy:}
\newline
The comparison of the trajectories for \snr1993j 
for the four possibilities, classic or  relativistic,
conservation of energy or momentum, 
is presented in Table \ref{tablecomparison}. 
\begin{table}
\begin{center}
\caption
{
Type of regime, conservation,
model, $\chi^2$ and reference
for \snr1993j 
}
\begin{tabular}{|l| l| l| l|l|}
\hline
regime  & conservation 
& model  & $\chi^2$      &  Reference  \\ \hline
classical &momentum  &  
inverse~ power ~ law  & 
276   & 
Figure~6~in~ \cite{Zaninetti2011a}
\\ \hline
classical &momentum & Plummer ~profile & 265 & 
Figure~8~in~ 
\cite{Zaninetti2014f}
\\ \hline
relativistic & momentum  &  
Lane--Emden ~ profile  & 
471   & 
Figure~10~in~ 
\cite{Zaninetti2014f}
\\ \hline
relativistic  & energy &  
power~law ~ profile  & 
6387  & 
Figure~4~in~ \cite{Zaninetti2020c} 
\\ \hline
relativistic  & energy &  
exponential ~ profile  & 
13145  & 
Figure~6~in~ \cite{Zaninetti2020c} 
\\ \hline
relativistic  &energy &  
Emden ~ profile  & 
8888  & 
Figure~8~in~ 
\cite{Zaninetti2020c} 
\\ \hline
classical & energy &  
power~law ~ profile  & 
176.6   & 
Figure~ \ref{trajpowerlaw}~in~  this~paper 
\\ \hline
relativistic &energy &  
NFW~profile  & 
823   & 
Figure~ \ref{nfwrelsn1993j}~in~ 
this~paper 
\\ \hline
relativistic & energy &  
NCD  & 
9589   &
Figure~ \ref{ncdrelsn1993j}~in~ 
this~paper 
\\ \hline
 \end {tabular}
\label{tablecomparison}
\end{center}
\end{table}

The best results  are obtained for the energy conservation
in the presence of a   power law profile in the 
present paper, see equation (\ref{traiettoriapotenza}).

{\bf Light curve:}
\newline  
The luminosity in the various  astronomical
bands is here assumed to be proportional
to  the classical or relativistic
flux  of mechanical kinetic energy.
This theoretical dependence  is not enough
and the concept of optical depth
should be introduced.
Due to the complexity of the time dependence 
of  the optical depth, a polynomial approximation
of degree $M$ with time as independent variable  
has been suggested, 
see equation~(\ref{polynomial}) which is used in a linear
or logarithmic form.
\newline
{\bf Comparison with astronomical data:}
\newline  
The framework  of conversion   of
the classical flux  of mechanical kinetic energy
into the various astronomical  bands coupled
with a time dependence  for the optical depth
allowed simulating the  various morphologies of the
LC of supernovae.
In particular, in the case of \snr1993j
we modeled:
(i) the $R$ LC of \snr1993j over 10 yr,
see Figure  \ref{mtpowerlaw}, 
(ii)   
the  $H-\alpha$ 
and the  2.0--8.0 keV luminosities  over 10 yr, 
see Figure \ref{ltpowerlaw},
(iii)
the radio flux density
over 443 days,
see Figure \ref{radiocomptaunu}
and 
(iv)
$V$ LC over 63 days,
see the double peak 
visible in Figure \ref{duemagfew}.
The LC of  
of \grb
at 0.2--10 keV  
was modeled in Figure \ref{obs_theo_grb}
and that   of UVOT (U)
+ HST (F330W) for \rgb2
was modeled in Figure \ref{ncd_theo_obs_rgb}.
\newline
{\bf Magnetic field}
\newline  
The 
minimum magnetic field depends on the luminosity 
and this  allows to derive  its theoretical 
dependence on time, 
see Figure \ref{magneticsn1993j}.
The above dependence allows deriving  the  
distance for the 
mean free path between accelerating clouds
for the  Fermi II mechanism
when the relativistic electron emits
synchrotron radiation 
in the keV region, 
see Figure \ref{linequality}.
\providecommand{\newblock}{}

\end{document}